# Optimality of the genetic code with respect to protein stability and amino acid frequencies

Dimitri Gilis[1], Serge Massar[2], Nicolas Cerf[3], and Marianne Rooman[1,2]

1. Biomolecular Engineering, Université Libre de Bruxelles, CP 165/64, 50 av. F. D. Roosevelt, 1050 Bruxelles, Belgium

2. Service de Physique Théorique, Université Libre de Bruxelles, CP225, bld. du Triomphe, 1050 Bruxelles, Belgium

3. Ecole Polytechnique, Université Libre de Bruxelles, CP 165/56, 50 av. F. D. Roosevelt, 1050 Bruxelles, Belgium

## Abstract

How robust is the natural genetic code with respect to mistranslation errors? It has long been known that the genetic code is very efficient in limiting the effect of point mutation. A misread codon will commonly code either for the same amino acid or for a similar one in terms of its biochemical properties, so the structure and function of the coded protein remain relatively unaltered. Previous studies have attempted to address this question more quantitatively, namely by statistically estimating the fraction of randomly generated codes that do better than the genetic code regarding its overall robustness. In this paper, we extend these results by investigating the role of amino acid frequencies in the optimality of the genetic code. When measuring the relative fitness of the natural code with respect to a random code, it is indeed natural to assume that a translation error affecting a frequent amino acid is less favorable than that of a rare one, at equal mutation cost (being measured, e.g., as the change in hydrophobicity). We find that taking the amino acid frequency into account accordingly decreases the fraction of random codes that beat the natural code, making the latter comparatively even more robust. This effect is particularly pronounced when more refined measures of the amino acid substitution cost are used than hydrophobicity. To show this, we devise a new cost function by evaluating with computer experiments the change in folding free energy caused by all possible single-site mutations in a set of known protein structures. With this cost function, we estimate that of the order of one random code out of 100 millions is more fit than the natural code when taking amino acid frequencies into account. The genetic code seems therefore structured so as to minimize the consequences of translation errors on the 3D structure and stability of proteins.

## Introduction

One of the tantalizing questions raised by molecular biology is whether the basic structures of life as we know them arose through a Darwinian evolutionary process, and if so, what were the evolutionary pressures acting on them? One such structure which could have changed through evolution is the genetic code. The genetic code is almost universal throughout life, with very minor variations in mitochondria, and any trace of the different stages of a possible evolutionary process, if any, has disappeared. Nevertheless the idea that the genetic code could have evolved to its present form has been repeatedly suggested in the literature (1). For instance, it has been proposed that early codes were simpler in that they coded for only a few amino acids, and that the number of amino acids coded in the genetic code increased as the code evolved (2-5). Several hypotheses have been put forward to explain the evolution of the genetic code to its present form, and to find out what is the genetic code optimized for (4, 6-15). One possible scenario is that the genetic code evolved so as to minimize the consequence of errors during transcription and translation (7, 8, 10, 11, 16). To test this hypothesis, some authors tried to estimate the percentage of achievement of the natural code by quantifying the cost of single-base changes (17-19).

More recently, Haig and Hurst (20) and Freeland and Hurst (21) improved the latter approach by comparing the natural code with random codes. To this end, they defined a fitness function $\Phi$ that measures the efficiency of the code to limit the consequences of transcription and translation errors. This function $\Phi$ supposedly evolved towards a minimum through evolution. To measure how close the natural code is to the actual minimum of $\Phi$, they generated random genetic codes, and computed the fraction of those that are better - i.e., have a smaller value of $\Phi$ - than the natural code. They found that only a very small fraction of the random codes are better than the natural code, and concluded that the natural code is therefore optimal in that it minimizes the effect of translation and transcription errors.

Haig and Hurst (20) tested several fitness functions Φ based on different physico-chemical parameters, and found that single-base changes in the natural code had the smallest average effect when using, as a cost measure, the change in polarity or hydropathy between the corresponding amino acids. These parameters, although not unique, are clearly biologically relevant as they are related to hydrophobicity, a property known to play an important role in protein conformation. Changing, through a transcription or a translation error, a non-polar amino acid into a polar one at some strategic position in the sequence of a protein can have dramatic consequences on its conformation. Using these parameters, and assuming that all point mutations occur with the same frequency, Haig and Hurst (20) found that the fraction of random codes that beat the natural code is of the order of $10^{-4}$.

It has been shown experimentally that individual translation errors occur more frequently at the first and third codon positions than at the second (8, 22, 23), and that there are transition/transversion biases (24-27). Taking this into account, Freeland and Hurst (21) proposed a modified fitness function Φ which models more accurately the probability of translation errors. They found that with this improved modeling, the fraction of random genetic codes that are better than the natural one decreases from $10^{-4}$ to $10^{-6}$. They retrieved from their calculations a well known property of the genetic code: single-base substitutions in the first and third codon position are strongly conservative with respect to changes in polarity (8, 28).

In the present paper*, we highlight the importance of another parameter in the optimization of the genetic code, namely the frequency with which different amino acids occur in proteins. This frequency differs from protein to protein, and even from species to species, but there is a general pattern that prevails (Table 1). For instance Leu is the most common amino acid, and Trp the rarest. In Fig. 1, we have plotted the number of codons coding for the same amino acid (synonyms) versus the amino acid frequency. The clear correlation between these two quantities, first noted by King and Jukes (30), led us to suspect that the amino acid frequency is an important parameter in the optimization of the genetic code, which should also be taken into account in the fitness function $\Phi$. Our calculations indeed confirm that the genetic code is even more optimal with respect to translation errors if the amino acid frequencies of Table 1 are properly incorporated in $\Phi$.

---

* These results have already been presented at the Conference *Frontiers of Life,* Blois, France (June 2000).

In addition, we bring further improvements to the fitness function $\Phi$ by using other quantities than polarity to measure the roles of the different amino acids in protein conformation and stability. It should be stressed that the biological relevance of the parameters used in $\Phi$ is crucial in the estimation of the relative robustness of the natural code. Indeed, one can always construct an artificial fitness function $\Phi$ such that the natural biological structure apparently lies at its minimum. Clearly, the hydrophobicity parameters used by Haig and Hurst (20) are biologically motivated, but we would like to do better by refining our cost measure. In particular, we devise a mutation matrix describing the average cost of single amino acid substitutions in protein stability, obtained by computer experiments. This mutation matrix combines many different physico-chemical properties of the amino acids. For instance, it takes into account that mutating Cys into any other amino acid is always very costly since this can break a disulfide bond. Such an effect would not be apparent if only a single property, say hydrophobicity, was taken into account. We show that, with a fitness function $\Phi$ depending on this mutation matrix and the amino acid frequencies, only about 1 out of $10^8$ randomly generated codes are better than the natural code. This suggests that the genetic code is even better optimized to limit translation errors than was previously thought. After completion of this paper, we became aware that Freeland et al. (29) have recently improved the fitness function in a different way by using as a cost measure an amino acid substitution matrix derived from protein sequence alignments. They found that the relative fitness of the genetic code increases with this more realistic cost function. It is reassuring that this conclusion is reached using both our mutation matrix and Freeland et al.'s substitution matrix, although the derivation of the two cost functions use very different starting points (protein 3D structures in one case, sequence similarity in the other).

## Fitness of the genetic code with respect to translation errors

Consider the natural genetic code. It is built out of 64 codons, each consisting of 3 consecutive DNA bases (A, G, C, T) or RNA bases (A, G, C, U). These 64 codons are divided into 21 sets of synonyms, which code each for one of the 20 natural amino acids or correspond to a stop signal; hence, to each codon *c*, an amino acid (or stop signal) *a* is assigned through a function *a(c)*. Consider now an error during transcription from DNA to RNA or during translation from RNA to protein, in which codon *c* is mistaken for codon *c'*. This error thus results in amino acid *a(c)* being replaced by amino acid *a'=a(c')*. The associated cost is estimated by a function *g(a,a')*, which measures the difference between the amino acids *a* and *a'* with respect to their physico-chemical properties or their role in (de)stabilizing protein structures; when *a* or *a'* corresponds to a stop codon, we set *g(a,a')=0*. Different cost functions *g* will be discussed in the next section. Following Freeland and Hurst (21), the fitness Φ of a code is measured by the average of the cost *g* over all codons *c* and all single-base errors *c→c'*:

$$\Phi^{FH} = \frac{1}{64} \sum_{c=1}^{64} \sum_{c'=1}^{64} p(c'|c)\, g(a(c),a(c')) \quad , \tag{1}$$

where *p(c'|c)* is the probability to misread codon *c* as codon *c'*. If one focuses on transcription errors only, as Haig and Hurst (20), then all *p(c'|c)'s* must be taken equal. But here we consider translation errors, as Freeland and Hurst (21), and hence *p(c'|c)* changes according to whether *c* and *c'* differ in the first, second or third base, and lead to a transition or a transversion. A transition is the substitution of a purine (A, G) into another purine, or a pyrimidine (C, U/T) into another pyrimidine, whereas a tranversion interchanges purines and pyrimidines. Based on experimental data indicating that transitions

are more common than transversions (24-27), and that errors on the third base are more frequent than errors on the first base, which are themselves more frequent than errors on the second base (8, 22, 23), Freeland and Hurst (21) have chosen the following values of $p(c'|c)$, which we also use here:

$p(c'|c) = 1/N$    if c and c′ differ in the 3$^{rd}$ base only,

$p(c'|c) = 1/N$    if c and c′ differ in the 1$^{st}$ base only and cause a transition,

$p(c'|c) = 0.5/N$   if c and c′ differ in the 1$^{st}$ base only and cause a transversion,

$p(c'|c) = 0.5/N$   if c and c′ differ in the 2$^{st}$ base only and cause a transition,

$p(c'|c) = 0.1/N$   if c and c′ differ in the 2$^{st}$ base only and cause a transversion,

$p(c'|c) = 0$        if c and c′ differ by more than 1 base,

where N is a normalization factor ensuring that $\Sigma_{c'} p(c'|c)=1$.

## Incorporating amino acid frequencies in the fitness function

Let us now come back to the correlations between the number of codons coding for an amino acid and the frequency of this amino acid (see Fig. 1). King and Jukes (30), who first noted this correlation, suggested that most of the amino acids in the genomes have arisen by random mutations which do not affect the properties and function of the proteins. As a consequence, the number of synonymous codons determines the frequency of amino acids. The fitness function $\Phi^{FH}$ is in accordance with this point of view, since each codon is given equal weight in this function.

An alternative interpretation, assuming a very different chain of causality, is that the amino acid frequencies are fixed by their physico-chemical properties. For instance, Trp would be a rare amino acid because its specific properties are seldom needed in proteins or because it is difficult to synthesize. The correlation between the amino acid frequencies and number of synonymous codons (Fig. 1) would then be interpreted as being due to an adjustment of the natural genetic code to the frequency of the amino acids. The conclusions which are reached using these two opposite interpretations are discussed in the final section of this paper.

A codon error substituting a rare amino acid into another has obviously less consequence, at least on the average, than an error affecting a frequent amino acid. The frequencies with which the different amino acids occur in proteins, which are approximately universal in all organisms (Table 1), are only imperfectly taken into account in the fitness function $\Phi^{FH}$ given by eq. (1), because of the imperfect correlation between amino acid frequency and number of synonymous codons (Fig. 1). In order to properly account for this effect, we propose a modified fitness function $\Phi^{faa}$:

$$\Phi^{faa} = \sum_{c=1}^{64} \frac{p(a(c))}{n(c)} \sum_{c'=1}^{64} p(c'|c)\, g(a(c),a(c')) \quad , \tag{2}$$

where $p(a)$ is the relative frequency of amino acid $a$, and $n(c)$ is the number of codons in the block to which $c$ belongs. In other words, $n(c)$ is the number of synonyms coding for the amino acid $a(c)$ that $c$ codes for. Note that eq. (2) supposes that there is no codon bias, i.e., the different synonyms of a given amino acid appear with the same frequency.

In order to measure the effect of the amino acid frequency on the value of the fitness function $\Phi^{faa}$, we define, for the sake of comparison, another fitness function $\Phi^{equif}$ where all the amino acids are supposed equifrequent, i.e. $p(a)=1/20$:

$$\Phi^{equif} = \frac{1}{20} \sum_{c=1}^{64} \frac{1}{n(c)} \sum_{c'=1}^{64} p(c'|c)\, g(a(c),a(c')) \quad . \tag{3}$$

## Cost of substituting an amino acid into another

The function *g(a,a')* in eqs (1) and (2) measures the cost - as far as protein stability and structure is concerned - of substituting amino acid a by a'. This cost depends on several physico-chemical and energetic factors. Hydrophobic interactions are known to constitute the dominating energetic contribution to protein stability. Hence, a natural choice for *g* consists of taking the squared difference in hydrophobicity *h* of the amino acids *a* and *a'*:

$$g^{hydro}(a,a') = \left(h(a) - h(a')\right)^2 \qquad . \qquad (4)$$

There exist various hydrophobicity scales for amino acids. We have tested two of them. The first is the polarity scale defined by Woese et al. (31), which is the one that was used by Haig and Hurst (20) and Freeland and Hurst (21). In the second scale, *h(a)* is the average solvent accessibility of amino acid *a* derived from a set of 141 well resolved and refined protein structures with low sequence identity (see Appendix); solvent accessibilities are computed using SurVol (32). We denote the associated cost functions as $g^{pol}$ and $g^{access}$, respectively.

Although hydrophobic forces dominate in proteins, other types of interactions also contribute to protein stability. We therefore also attempted to devise a better cost function *g(a,a')*, measuring more accurately the difference between amino acids *a* and *a'*. This new function is inspired by recent computations of the change in free energy of a protein when a single amino acid is mutated (33-35). It is obtained by mutating *in silico*, in all proteins of the aforementioned set of 141 protein structures, and at all positions, the wild type amino acids into the 19 other possible ones, and evaluating the resulting changes in folding free energy with mean force potentials derived from the same structure dataset. The matrix elements *M(a,a')* are obtained as the average of all the computed folding free energy changes which correspond to a substitution *a→a'*. Details on the procedure and the value of the matrix elements *M(a,a')* are given in the Appendix. This matrix is taken as a cost function:

$$g^{mutate}(a,a') = M(a,a') \qquad . \tag{5}$$

As a last cost function, we consider the "blosum62" substitution matrix (36, 37), one of the most commonly used matrices in the context of protein sequence alignment:

$$g^{blosum}(a,a') = blosum62(a,a') \qquad . \tag{6}$$

This matrix is computed from the frequency of amino acid substitutions in families of evolutionary related proteins. However, it reflects not only the similarity between amino acids with respect to their physico-chemical and energetic properties, but also the facility with which one amino acid is mutated into another and thus their proximity in the genetic code. Strictly speaking, it should therefore not be used to estimate the fitness of the genetic code; we only use it here as a reference. This potential problem might also affect the substitution matrix used by Freeland et al. (29), but probably to a smaller extent as their matrix was derived from highly diverged protein sequences.

## Results: the genetic code versus random codes

To evaluate the robustness of the natural genetic code with respect to translation errors, we computed the fitness functions $\Phi^{FH}$, $\Phi^{equif}$ and $\Phi^{faa}$ using eqs (1-3) for the natural genetic code, and compared it to the corresponding fitnesses of random codes. The random codes are obtained by maintaining the codon block structure of the natural genetic code, where each block corresponds to synonyms coding for the same amino acid (or stop signal). When generating a random code, the stop signal is kept assigned to the same block as in the natural genetic code, whereas the different amino acids are randomly interchanged among the 20 remaining blocks. Thus, each random code is simply specified by a different function $a(c)$ in eqs (1-3). This is the procedure previously used by Haig and Hurst (20) and Freeland and Hurst (21).

Thus, in a first stage, we computed the fitness functions $\Phi^{equif}$ and $\Phi^{faa}$ for the natural genetic code and for $10^8$ randomly generated codes, using the three cost functions $g^{pol}$, $g^{access}$ and $g^{mutate}$.

We then calculated the fraction *f* of random codes whose value of Φ is lower than that of the natural code. This fraction is supposedly a good estimate of the relative merit of the natural genetic code comparatively to other codes. The results are given in Table II. It appears that, for all cost functions *g*, this fraction *f* is between 10 and 100 times smaller for $\Phi^{faa}$ than for $\Phi^{equif}$. This indicates that the natural code appears to be better optimized with respect to translation errors if the amino acid frequencies are taken into account.

In order to investigate this further, we have analyzed which of the cost functions $g^{pol}$, $g^{access}$ or $g^{mutate}$ the genetic code appears to be best optimized for. For this purpose, we compared the fraction *f* of better codes for each of the cost functions using the fitness function $\Phi^{faa}$ (cf. Table II). For the hydrophobicity functions $g^{pol}$ and $g^{access}$, the result is roughly the same: *f* is about 1-8 in a million. The relative statistical error on this value is of the order of $N^{-1/2}$, where *N* is the number of random codes better than the natural one that were found in our sample of $10^8$ random codes; thus, *N* is about 100-800, and the error is insignificant. For the mutational cost function $g^{mutate}$, we did not find any random code better than the natural one among the $10^8$ random codes. Then, to estimate the fraction *f* without having to generate a larger ensemble, we used the following procedure. We computed, from the values of $\Phi^{faa}$ for the $10^8$ random codes, the probability function $\pi(\Phi^{faa})$ to have a given value of $\Phi^{faa}$. We fitted $log(\pi(\Phi^{faa}))$ to a polynomial of fourth degree, and extrapolated this curve down to the value of $\Phi^{faa}$ for the natural code. This provides an estimate of the fraction *f* of random codes that have a lower $\Phi^{faa}$ value. Note that this estimate is essentially insensitive to the degree of the polynomial. We found that using $g^{mutate}$ as a cost function this fraction is of the order of 1 in $10^8$.

This result shows that the natural genetic code appears even more optimal if the cost function $g^{mutate}$ is used than if hydrophobicity-based cost functions are considered. Since $g^{mutate}$ has been computed from protein stability changes effected by point mutations, we may conclude that evolution has optimized the genetic code in such a way as to limit the effect of translation errors on the 3D structure and stability of the coded proteins. Note that the improvement brought by the choice of $g^{mutate}$ results from the fact that it probably better accounts for the cost of a mutation than a mere difference of hydrophobicity; for example, Gly, Pro, and Cys have close neighbors in hydrophobicity, while the cost of their mutation as accounted for by $g^{mutate}$ is high. This is due to their special role in determining protein structure: Gly and Pro can adopt backbone torsion angles essentially inaccessible to other amino acids, and Cys can form disulfide bonds.

For completeness, we have added in Table II the values of the fraction $f$ of random codes with a lower $\Phi^{faa}$ value than the natural one, using the cost function $g^{blosum}$. With this function, $f$ (extrapolated as above) is about three times smaller than with $g^{mutate}$. This was expected as the blosum matrix (36, 37) is computed from amino acid substitutions in families of evolutionary related proteins, which are more frequent between amino acids that are closer in the genetic code. Using $g^{blosum}$ can therefore be considered as superimposing some information on the proximity of amino acids in the genetic code to the desired measure of their similarity in preserving protein structure. So, it is not surprising that $g^{blosum}$ does better in minimizing $\Phi^{faa}$ than $g^{mutate}$, which only includes information related to protein structure. In contrast with what happens with $\Phi^{faa}$, the fraction of random codes having a lower $\Phi^{equif}$ value than the natural code is *larger* when using $g^{blosum}$ than with $g^{mutate}$. Thus, if all amino acids are assumed to be equifrequent then the apparent merit of $g^{blosum}$ disappears. This means that, besides informations about proximity, $g^{blosum}$ also incorporates information about the amino acid frequencies. For these reasons, $g^{blosum}$ is probably an intrinsically bad cost measure for our purposes here.

Finally, we have attempted to check the significance of our main result that the natural code is better optimized if amino acid frequencies are taken into account. To this end, we have computed the fraction $f$ of random codes that beat the natural one for *random* choices of the amino acid frequencies, distinct from the natural frequencies $p(a)$. We have generated $10^2$ sets of random $p(a)$'s, and, for each of them, estimated the fraction *f* (out of a sample of $10^6$ random codes). The percentage of random amino acid frequency sets that result in a lower fraction *f* than the natural frequencies is shown in Table III. We find that a random assignment of the amino acid frequencies does not decrease *f* in the great majority (at least 94 %) of the cases, and this tendency persists for all cost functions *g*. Thus, the probability that the decrease of *f*, observed in Table II, when passing from $\Phi^{equif}$ to $\Phi^{faa}$, was due to chance is quite limited. We may therefore conclude that the genetic code is optimized so as to take into account the *natural* amino acid frequencies.

For comparison, we have also included in Table II the results based on the fitness function $\Phi^{FH}$. It can be argued that this function takes in part, but imperfectly, the amino acid frequencies into account. Indeed, for this fitness function each codon is assigned the same weight, which corresponds to each amino acid being assigned a frequency proportional to the number of synonyms *n(a)* coding for it. In the case of the natural genetic code, this frequency corresponds approximatively to the amino acid frequency since there is a correlation between *n(a)* and *p(a)*, as shown in Fig. 1. But for random codes, where the amino acids are randomly interchanged between the codon blocks, this correspondence breaks down. Thus, the way in which $\Phi^{FH}$ takes amino acid frequencies into account depends on the code considered. This explains why the fraction *f* of random codes better than the natural one is sometimes smaller or sometimes larger using $\Phi^{FH}$ instead of $\Phi^{equif}$. Note, however, that *f* is always larger for $\Phi^{FH}$ than for $\Phi^{faa}$, indicating again the importance of the amino acid frequencies in the optimality of the genetic code.

## Discussion and conclusion

Our results confirm and specify those of Freeland and Hurst (21): the genetic code seems structured so as to minimize the consequences of translation errors on the 3D structure and stability of the coded proteins. We have shown that, using the cost function $g^{mutate}$, which best reflects the roles of various amino acids in protein structures, and taking amino acid frequencies into account, about 1 out of $10^8$ random codes does better than the natural code. However, we have to keep in mind that there exist $20! \approx 2 \cdot 10^{18}$ possible codes preserving the codon block structure, which means that we can expect about $10^{10}$ better codes overall. Moreover, if the codon block structure is not preserved (14), the number of possible codes is larger by orders of magnitude, and therefore the number of codes better than the natural one will certainly be much larger.

So, we can assert from our analysis that the genetic code has been optimized through evolution up to a certain point, even though it is probably not fully optimal at least with respect to the parameters considered here. However, our analysis does not give us information about the mechanism of this evolution since there is unfortunately no trace left of evolution of the code or amino acid frequencies in early times. For instance, we do not know whether the relative frequency of occurrence of amino acids in proteins adapted so as to increase the optimality of the genetic code with respect to translation errors, or, on the contrary, whether the genetic code evolved to take into account pre-existing amino acid frequencies. We can, however, argue that if the amino acid frequencies adapted to the genetic code, as assumed by King and Jukes (30), a discrepancy in amino acid composition between frequently and unfrequently expressed genes might be detectable today (unless the period during which evolution took place was long enough for this discrepancy to vanish). If, alternatively, the genetic code adapted to the amino acid frequencies, and thus if these frequencies acted as an evolutionary pressure, one can imagine two scenarios. Either the code optimized to take into account the pre-biotic frequencies of the amino acids that became involved in it, or it optimized for the amino acid frequencies of already formed proteins (or of a subset of them) that were important for life and maybe linked to the code's control. Perhaps can we assume, more realistically, that the genetic code and amino acid frequencies *coevolved* during some evolutionary period, thereby approaching an optimal code/amino acid relation.

More generally, the parameters that acted as evolutionary pressure on the genetic code probably included all the mechanisms that code and maintain the genetic information, and were not just restricted to the frequency of amino acids and the preservation of protein structure. For example, the genetic code is obviously related to the translation apparatus, composed of the ribosomes and transfer RNA, whose action we described schematically here by the probabilities $p(c'/c)$ to misread codon $c$ as $c'$. This apparatus was certainly less reliable at the beginning of evolution. All these mechanisms probably evolved together with the genetic code during the early stages of life.

The evolution of the code came to an end at an early stage of life development, as reflected by its universality among all organisms. This probably arose because even small modifications in the code would entail loss of functionality of previously expressed genes. Moreover, the advent of more sophisticated transcription/translation control mechanisms, which involve huge protein systems, could have decreased the evolutionary pressure on the genetic code. Even though the present data on the genetic code are insufficient to discriminate between evolution scenarios, our analysis enables us to put some constraints on the situation at the time when the code evolution was frozen. In particular, it appears that the frequencies of the amino acids that were used in proteins synthesized at that time were similar to the present frequencies. We do not know what determines the present amino acid frequencies, but, presumably, they are due at least in part to their physico-chemical properties. For instance, the hydrophobic to hydrophylic ratio is intrinsically related to the globular structure of proteins and certainly contributes to the pressure on amino acid frequencies. Also, amino acid that are easily synthesized may be used more often. Thus, we can assert that some of the pressures that determine the present amino acid frequencies were already present at the time when the code took its definitive form. In addition, the increased optimality of the genetic code with respect to $g^{mutate}$ implies that the 3D structure of proteins probably played an equally important role in fixing the structure of the code. Since the 3D structure of a protein essentially determines its function, this suggests, more generally, that the protein function acted as a main evolutionary pressure on the code structure. Consequently, at the time when the genetic code took its present form, primitive life was presumably synthesizing complex proteins already. This provides a tentative picture of primitive life at that time: the translation apparatus was similar to the present one, and organisms where made of complex proteins whose amino acid frequencies was comparable to the present ones.

## Acknowledgments

We are grateful to Jacques Reisse for fruitful discussions and to a referee for useful comments. DG benefits from a "FIRST-université" grant of the Walloon Region. SM and MR are, respectively, research associate and senior research associate of the Belgian National Fund for Scientific Research.

## Appendix: Derivation of the mutation matrix

The derivation is based on a dataset of 141 high-resolution protein structures determined by X-ray crystallography listed in Wintjens et al. (38). In order to avoid biases, these 141 proteins are chosen so as to either present less than 20% sequence identity or to present less than 25% sequence identity and no structure similarity.

The protein main chains are described by their heavy atoms, and each side chain is represented by a pseudo-atom $C^\mu$. For a given amino acid type, the $C^\mu$ has a well-defined position relative to the main chain, corresponding to the geometric average of all heavy side chain atoms of this type in the dataset (39); for glycine, the $C^\mu$ pseudo-atom is positioned on the $C^\alpha$. Side chain degrees of freedom are thus neglected.

Each residue, at each position of each of the 141 proteins, is mutated in turn into the 19 non-wild type amino acids. The mutations are performed by keeping the main chain structure unchanged, and substituting the $C^\mu$ of the mutated amino acid by that of the mutant amino acid. For each of these mutations, the change in folding free energy is evaluated using database derived potentials. For each substitution of amino acid $a$ into $a'$, the average of all computed changes in folding free energy, at all protein positions, is computed and defined as minus the matrix element $M(a,a')$. We then symmetrize $M$ by setting $M(a,a')=[M(a,a')+M(a',a)]/2$ and only consider the lower half of $M$ ($a \leq a'$). This procedure does not define the diagonal elements of $M$. Based on the principle that the structural role of a given amino acid is fulfilled by no other amino acid better than by itself, we assign to all the diagonal element the same maximum value: $M(a,a)= Max[M(a',a'')]+1$. Then, to simplify $M$ without modifying its structure, we center it around its mean value:

$$M(a,a') \rightarrow M(a,a') - <M> \qquad with \qquad <M> = \frac{1}{210} \sum_{a' \leq a} M(a,a')$$

Finally, we multiply all matrix elements *M(a,a')* by 2 and replace them by the closest integer. The resulting half matrix is given in Fig. 2. For more details, see the supplementary material.

| Amino acid | p(a) (%) |
|---|---|
| Ala | 7.85 |
| Arg | 5.33 |
| Asp | 5.37 |
| Asn | 4.55 |
| Cys | 1.88 |
| Glu | 5.83 |
| Gln | 3.77 |
| Gly | 7.35 |
| His | 2.35 |
| Ile | 5.80 |
| Leu | 9.43 |
| Lys | 5.88 |
| Met | 2.28 |
| Phe | 4.07 |
| Pro | 4.56 |
| Ser | 6.04 |
| Thr | 6.17 |
| Trp | 1.31 |
| Tyr | 3.27 |
| Val | 6.92 |

Table I: The mean frequency *p(a)* with which the different amino acids *a* appear in the genomes of many different organisms, derived from the Swiss-Prot database (40) (see http://cbrg.inf.ethz.ch/ServerBooklet/section2_11.html).

| $f$ | $\Phi^{FH}$ | $\Phi^{equif}$ | $\Phi^{faa}$ |
|---|---|---|---|
| $g^{pol}$ | $2.7 \cdot 10^{-6}$ | $1.9 \cdot 10^{-5}$ | $1.5 \cdot 10^{-6}$ |
| $g^{access}$ | $3.3 \cdot 10^{-5}$ | $6.2 \cdot 10^{-5}$ | $8.4 \cdot 10^{-6}$ |
| $g^{mutate}$ | $1.8 \cdot 10^{-4}$ | $1 \cdot 10^{-6}$ | $1 \cdot 10^{-8}$ * |
| $g^{blosum}$ | $1.5 \cdot 10^{-3}$ | $1.1 \cdot 10^{-4}$ | $3 \cdot 10^{-9}$ * |

Table II: Fraction $f$ of random codes that have a lower value of the fitness function ($\Phi^{FH}$, $\Phi^{equif}$, or $\Phi^{faa}$) than the natural code, using each of the four cost functions $g^{pol}$, $g^{access}$, $g^{mutate}$ and $g^{blosum}$. The values with '*' have been obtained by extrapolation as explained in text.

|  | % |
|---|---|
| $g^{pol}$ | 6 |
| $g^{access}$ | 4 |
| $g^{mutate}$ | <1 |
| $g^{blosum}$ | <1 |

Table III: Percentage of the sets of random amino-acid frequency assignments for which the fraction $f$ of random codes that beat the natural code is lower than the corresponding fraction computed with the natural frequencies p(a)'s. This percentage is estimated using the four cost functions $g^{pol}$, $g^{access}$, $g^{mutate}$ and $g^{blosum}$. In the case of $g^{mutate}$ and $g^{blosum}$ we are only able to give an upper bound, because our sample of 100 random frequencies and $10^6$ random codes is too small. For these cost functions we found respectively 4 and 2 random frequencies for which none of the random codes was better than the natural code. The extrapolation method used for the natural frequencies showed that for none of these 6 frequency sets, the fraction $f$ was smaller than the corresponding fraction for the natural amino acid frequencies. Given the approximate character of this extrapolation, and the limited size of our sample, we give an upper bound of 1% for the fraction of random amino acid frequencies whose $f$ is smaller than for the natural frequencies.

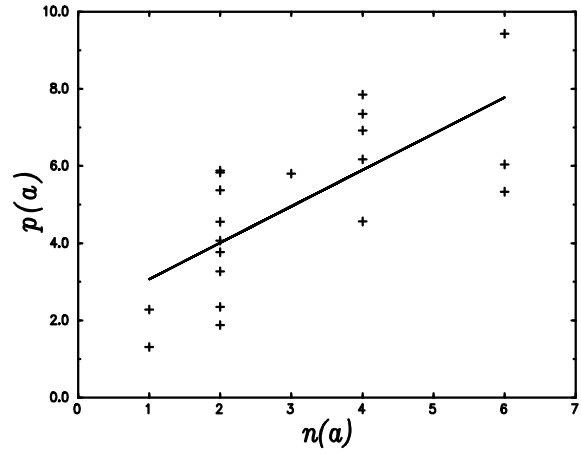

Figure 1: The relative frequency $p(a)$ (in %) of the amino acids $a$, taken from Table I, as a function of the number of synonyms $n(a)$ that code for them. The linear regression line is indicated; the correlation coefficient is equal to 0.71.

|   | A  | C  | D  | E  | F  | G  | H  | I  | K  | L  | M  | N  | P  | Q  | R  | S  | T  | V  | W  | Y  |
|---|----|----|----|----|----|----|----|----|----|----|----|----|----|----|----|----|----|----|----|----|
| A | +7 |    |    |    |    |    |    |    |    |    |    |    |    |    |    |    |    |    |    |    |
| C | -3 | +7 |    |    |    |    |    |    |    |    |    |    |    |    |    |    |    |    |    |    |
| D |  0 | -4 | +7 |    |    |    |    |    |    |    |    |    |    |    |    |    |    |    |    |    |
| E |  0 | -5 | +2 | +7 |    |    |    |    |    |    |    |    |    |    |    |    |    |    |    |    |
| F |  0 | -2 | -2 | -3 | +7 |    |    |    |    |    |    |    |    |    |    |    |    |    |    |    |
| G | -2 | -4 | -1 | -2 | -3 | +7 |    |    |    |    |    |    |    |    |    |    |    |    |    |    |
| H | +1 | -2 | +1 | +1 | -1 |  0 | +7 |    |    |    |    |    |    |    |    |    |    |    |    |    |
| I | -2 | -3 | -4 | -4 |  0 | -4 | -2 | +7 |    |    |    |    |    |    |    |    |    |    |    |    |
| K | +1 | -5 | +2 | +3 | -2 | -1 | +1 | -3 | +7 |    |    |    |    |    |    |    |    |    |    |    |
| L | -1 | -3 | -3 | -3 |  0 | -4 | -2 |  0 | -2 | +7 |    |    |    |    |    |    |    |    |    |    |
| M | +1 | -3 | -1 | -1 |  0 | -2 |  0 |  0 | -1 | +1 | +7 |    |    |    |    |    |    |    |    |    |
| N |  0 | -3 | +2 | +2 | -2 |  0 | +2 | -3 | +2 | -2 | -1 | +7 |    |    |    |    |    |    |    |    |
| P | -3 | -6 | -1 | -2 | -4 | -4 | -2 | -5 | -1 | -5 | -4 | -1 | +7 |    |    |    |    |    |    |    |
| Q | +1 | -4 | +2 | +3 | -1 | -1 | +2 | -2 | +3 | -1 |  0 | +2 | -2 | +7 |    |    |    |    |    |    |
| R | +1 | -4 | +2 | +2 | -2 | -1 | +1 | -3 | +3 | -2 |  0 | +2 | -2 | +2 | +7 |    |    |    |    |    |
| S | +1 | -2 | +2 | +1 | -1 |  0 | +2 | -2 | +2 | -1 |  0 | +2 | -1 | +2 | +2 | +7 |    |    |    |    |
| T |  0 | -2 | +1 |  0 |  0 | -1 | +2 | -1 | +1 | -1 |  0 | +1 | -1 | +1 | +1 | +2 | +7 |    |    |    |
| V | -1 | -3 | -3 | -4 |  0 | -4 | -2 | +1 | -3 |  0 |  0 | -3 | -4 | -2 | -2 | -1 |  0 | +7 |    |    |
| W | +1 | -2 |  0 | -1 |  0 | -2 | +1 | -1 |  0 | -1 | +1 |  0 | -3 |  0 |  0 | +1 |  0 | -1 | +7 |    |
| Y |  0 | -1 | -1 | -2 | +1 | -1 | +1 | -1 | -1 | -1 |  0 |  0 | -3 |  0 | -1 | +1 | +1 |  0 | +1 | +7 |

Figure 2: Mutation matrix *M*

## Supplementary material: details of the derivation of the mutation matrix

**Database derived potentials**

The potentials we use to evaluate the protein conformations are derived from observed frequencies of sequence and structure patterns in the aforementioned dataset of 141 proteins. We consider two types of potentials, called torsion (1, 2) and $C^\mu$-$C^\mu$ (3) potentials.

Torsion potentials describe only local interactions along the sequence. They take into account the propensities of single residues and residue pairs to be associated with a ($\varphi$, $\psi$, $\omega$) backbone torsion angle domain. Seven ($\varphi$, $\psi$, $\omega$) domains are considered, defined in Rooman et al. (1). We use two variants of the torsion potential, called torsion$_{\text{short-range}}$ and torsion$_{\text{middle-range}}$. Both are computed from propensities of a ($\varphi$, $\psi$, $\omega$) domain $t_i$, at position $i$ along the sequence, or pairs of domains ($t_i$, $t_j$), at positions $i$ and $j$, to be associated with an amino acid $a_k$ at position $k$. But we have $k-1 \leq i,j \leq k+1$ for the torsion$_{\text{short-range}}$ potential and $k-8 \leq i,j \leq k+8$ for the torsion$_{\text{middle-range}}$ potential. The folding free energy $\Delta G(S,C)$ of a sequence $S$ in the conformation $C$ computed from these propensities is expressed as (4, 5):

$$\Delta G_{torsion}(S,C) = -kT \sum_{i,j,k=1}^{N} \frac{1}{\zeta_k} \ln\frac{P(a_k,t_i,t_j)}{P(t_i,t_j)P(a_k)}$$

where $P$ are normalized frequencies, N is the number of residues in the sequence $S$, $k$ is the Boltzmann constant and $T$ is a conformational temperature taken to be room temperature (6). The normalization factor $\zeta_k$ ensures that the contribution to $\Delta G(S,C)$ of each residue in the window [$k-1$, $k+1$] for the torsion$_{\text{short-range}}$ potential or [$k-8$, $k+8$] for the torsion$_{\text{middle-range}}$ potential is counted once. It is equal to the window width, except near the chain ends.

The $C^\mu$-$C^\mu$ potentials are distance potentials dominated by non-local, hydrophobic interactions. They are based on propensities of pairs of amino acids ($a_i$,$a_j$) at position $i$ and $j$ along the sequence to be separated by a spatial distance $d_{ij}$, calculated between the pseudo atoms $C^\mu$. We consider two variants of $C^\mu$-$C^\mu$ potentials. The first one, called $C^\mu$-$C^\mu_{long\_range}$ potential, describes purely non-local interactions along the sequence, and only takes into account residues separated by at least 15 residues along the sequence, i.e. j≥i+16. The second one, simply called $C^\mu$-$C^\mu$ potential, though dominated by non-local interactions, possesses a local interaction component. The non-local component is obtained by considering together the frequencies of all residues separated by seven sequence positions and more, thus with j≥i+8. The local component is obtained by computing separately the frequencies of residues separated by one to six positions along the sequence, for i+1<j<i+8. Consecutive residues along the sequence are not considered. The folding free energies are expressed as:

$$\Delta G_{C^\mu-C^\mu}(S,C) = -kT \sum_{i<j}^{N} \ln \frac{P^{j-i}(a_i,a_j,d_{ij})}{P^{j-i}(a_i,a_j)P^{j-i}(d_{ij})}$$

with j≥i+16 and the normalized frequencies $P^{j-i}$ independent of $j-i$ for the $C^\mu$-$C^\mu_{long\_range}$ potential, and i+1<j and the normalized frequencies $P^{j-i}$ independent of $j-i$ for j≥i+8 for the $C^\mu$-$C^\mu$ potential. The discretisation of the spatial distances $d_{ij}$ is performed by dividing the distances between 3 and 8 Å into 25 bins of 0.2 Å width and merging the distances greater than 8 Å. To increase the reliability of the statistics, these bins are smoothed by combining the counts in each bin with those of the 10 flanking bins at each side. The predominance of the central bin is preserved by weighting the counts from each flanking bin by a factor 1/n, where n is the position relative to the central bin; n is equal to 1 for the two closest bins and to 10 for the two most distant bins.

The so-defined folding free energies are reliable for common amino acids and structure motifs, but not for less common ones. To correct for the sparse data, we substitute the sequence-specific frequencies $P(c,s)$, where $s$ denotes a sequence pattern and $c$ a structure motif, which appear in the two above equations defining the torsion and $C^\mu$-$C^\mu$ folding free energies, by a linear combination of these frequencies and the product of the separate frequencies of $s$ and $c$, denoted $P(s)$ and $P(c)$ respectively (7).

$$P(c,s) \rightarrow \frac{1}{\sigma+m^s}[\sigma P(c)P(s)+m^s P(c,s)]$$

where $m^s$ is the number of occurrences of the sequence pattern $s$ in the dataset, and $\sigma$ a parameter. This expression ensures that the sequence-specific contribution dominates for seldom sequence patterns and tends to zero for frequent ones. This behavior is modulated by the parameter $\sigma$, which we consider here equal to 50.

**Evaluation of folding free energy changes**

To estimate the stability changes caused by a single-site mutation, we compute the folding free energy changes as:

$$\Delta\Delta G(S_m,C_m;S_w,C_w) = \Delta G(S_m,C_m) - \Delta G(S_w,C_w)$$

where $C_m$ and $C_w$ are the mutant and wild-type conformations and $S_m$ and $S_w$ the mutant and wild-type sequences, respectively. With this convention, $\Delta\Delta G$ is positive when the mutation is destabilizing, and negative when it is stabilizing. The conformations $C_m$ and $C_w$ of the mutant and wild-type protein are assumed to be nearly identical. More precisely, the backbone conformations are taken as identical and only the position of the $C^\mu$ pseudo-atom, which is amino acid dependent, is different in the mutant and wild-type structures.

The folding free energies of the wild-type and mutant proteins are computed with linear combinations of the torsion and $C^\mu$-$C^\mu$ potentials described in the previous section. Previous analyses (8-10) have shown that the combination that gives the best evaluation of the $\Delta\Delta G$'s depends on the solvent accessibility $\mathcal{A}$ of the mutated residue; $\mathcal{A}$ is defined as the solvent accessible surface in the protein structure, computed by SurVol (11), times 100 and divided by its solvent accessible surface in an extended tripeptide Gly-X-Gly (12). These analyses have revealed that the mutations can be divided in three subsets. When the mutated residue is at the surface, with a solvent accessibility $\mathcal{A}$ equal to or larger than 50%, the optimal folding free energy changes has been shown to be equal to:

$$\Delta\Delta G_{\mathcal{A} \geq 50\%} = 1.14 \times \Delta\Delta G_{torsion_{short\_range}} + 0.27$$

When the mutated residue is half buried, half exposed to the solvent, with a solvent accessibility comprised between 20 and 40%, the optimal folding free energy is:

$$\Delta\Delta G_{20 < \mathcal{A} \leq 40\%} = 1.39 \times \Delta\Delta G_{torsion_{short\_range}} + 0.97 \times \Delta\Delta G_{C^{\mu}C^{\mu}} + 0.21$$

Finally, when the mutated residue is totally buried in the protein core, with a solvent accessibility less than or equal to 20%, the optimal folding free energy is:

$$\Delta\Delta G_{\mathcal{A} \leq 20\%} = 1.44 \times \Delta\Delta G_{torsion_{middle\_range}} + 1.70 \times \Delta\Delta G_{C^{\mu}C^{\mu}_{long-range}} + 1.44$$

When the mutated residue has a solvent accessibility comprised between 40 and 50%, we do not evaluate its folding free energy. We have indeed observed that in this case, the solvent accessibility of the mutated residue is not a good measure to guide the choice of the optimal potential.

## References of the supplementary material